# Ultra-Slow Dynamic Annealing of Neutron-induced Defects in n-type Silicon: Role of Charge Carriers

Ying Zhang[1,2], Yang Liu[1,2], Hang Zhou[1,2], Ping Yang[1,2], Jie Zhao[1,2], and Yu Song[1,2,†,*]

[1] Microsystem and Terahertz Research Center, China Academy of Engineering Physics, Chengdu 610200, China

[2] Institute of Electronic Engineering, China Academy of Engineering Physics, Mianyang 621999, China

**Abstract:** Neutron bombardments with equivalent fluence ($1\times10^{10}$ cm$^{-2}$) and different fluxes ($2.5\times10^{5}$ cm$^{-2}\cdot$s$^{-1}$ to $1\times10^{7}$ cm$^{-2}\cdot$s$^{-1}$) have been performed on three kinds of bipolar devices with n-type silicon as active regions. The measured increase of base currents and input bias currents are found to decrease with increasing neutron flux, implying that the strength of the dynamic annealing of divacancy defects in n-type silicon follows a positive flux dependence. Such a flux dependence is the same as that observed in ions implantation using protons, but the evident flux sensitivity in our experiment is 4 orders of magnitude lower than that of proton bombardment, despite the similarity in the masses and energies of the two particles. The huge discrepancy of flux range is attributed to the presence of vast charge carriers in proton bombardments, which strongly accelerate the dynamic annealing of defects by enhancing the diffusion velocity of Si interstitials and dissociation rate of defect clusters. Our work would contribute to the understanding of the defect annealing processes in silicon.

**Keywords:** Displacement damage, Dynamic Annealing, Neutron irradiation, BJT

† Present at College of Physics and Electronic Information Engineering, Neijiang Normal University, Neijiang 641112, China
*Corresponding author; E-mail: yusong@njtc.edu.cn

## I. INTRODUCTION

Bombardment of energetic particles induce atomic displacements and structural defects in crystalline semiconductors. The population of the created stable defects depends on the process of the defects' generation as well as their dynamic and thermal annealing during the bombardments. The dynamic annealing involves migration, recombination, and clustering of mobile point defects during irradiation [1, 2]. Factors of crucial importance for damage buildup in semiconductors are the energy, mass, fluence, and flux of the incident particles, as well as the temperature of the samples. Interestingly, the dependence of dynamic annealing on the fluence and flux show non-trivial characteristics. For the experiments of total fluence exceeding $1\times10^{12}$ cm$^{-2}$ and flux in the range of $1\times10^{10}\sim1\times10^{15}$ cm$^{-2}\cdot$s$^{-1}$, it is well recognized that, the strength of the dynamic annealing of defects decreases with increasing flux [3–8]. On the other hand, for the experiments of low total fluence of about $5\times10^{9}$ cm$^{-2}$ and low flux in the range of $1\times10^{7}\sim2\times10^{10}$ cm$^{-2}\cdot$s$^{-1}$, a reverse positive flux dependence of dynamic annealing is found in silicon bombarded by ions including proton [9], $^{4}$He [10], $^{11}$B, $^{12}$C, $^{16}$O, $^{28}$Si, $^{74}$Ge, $^{76}$Ge, and $^{120}$Sn [11–13]. In these experiments, the concentrations of the induced defects (such as $V_2^{2-}$, VO, $V_2^{-}$, and hydrogen-related defects) show sensitive dependence in certain flux range. It is remarkable that, the sensitive flux range becomes smaller for ions of heavier mass or larger energy [13]. For example, proton with the smallest mass shows the biggest sensitive flux ($1\times10^{10}$ cm$^{-2}\cdot$s$^{-1}$), while $^{120}$Sn with the biggest mass displays the smallest sensitive flux ($\sim2\times10^{8}$ cm$^{-2}\cdot$s$^{-1}$).

It should be noticed that, the incoming ions induce both non-ionizing and ionizing energy depositions in semiconductors, which can be measured by the non-ionizing energy loss (NIEL) and ionization energy loss (IEL) [14], respectively. Here, NIEL and IEL have close relationship with the nuclear and electronic stopping power well known by the ion beam community [15]. NIEL can be derived by inserting the function for nonionizing loss to the nuclear stopping power (Lindhard equation) which gives the fraction of transferred energy that is nonionizing [15, 16], while IEL takes into account the ionizing energy transfer [17, 18]. For 1MeV protons, the NIEL is $6.6×10^{-2}$ MeV cm$^2$/g and the IEL is $1.83×10^2$ MeV cm$^2$/g. The IEL/NIEL ratio is over $2.77×10^3$. Other ions also possess large portions of IEL in the total deposition energy. It is well known that, NIEL directly contributes to the construction of the displacement defects, while IEL generates electron-hole pairs. Previous research of injection annealing has identified that the presence of charge carriers can strongly enhance the dynamic annealing processes [19-23]. The origin comes from the enhanced mobility of isolated Si interstitials through alternating capture and release of electrons [24–28]. The IEL in the ion bombardments may generate higher concentrations of charge carriers than the injection, hence can more efficiently promote the dynamic annealing. How will the dynamic annealing behave if the vast background charge carriers were removed? Will the previously observed positive flux dependence still arise? To answer these interesting questions will not only be helpful to reveal the basic mechanism of the dynamic annealing but also be helpful to clarify the role of charge carriers in the defect buildup. However, to our best knowledge, there is rarely such investigation.

In this work, we investigate the flux dependence of the dynamic annealing in n-type silicon by neutron bombardment of PNP transistors. The neutrons are used instead of protons or other ions because they have much smaller IEL/NIEL ratio (~0.84) at an equivalent energy of 1 MeV [14]; the transistors are used instead of bulk silicon because the base currents are very sensitive to the low concentration defects. Equivalent fluence ($1×10^{10}$ cm$^{-2}$) neutron bombardments of PNP transistors and operational amplifiers are carried out in the flux range between $2.5×10^5$ cm$^{-2}$·s$^{-1}$ to $1×10^7$ cm$^{-2}$·s$^{-1}$ at room temperature. The measured increase of base currents and input bias currents are found to decrease with increasing flux. This fact implies that the positive flux effect of dynamic annealing also arises for neutron irradiation. However, it is found that the sensitive flux range in our neutron irradiation experiments is nearly 4 orders of magnitude lower than the sensitive flux of proton bombardment experiments. In other words, ultra-slow dynamic annealing happens for neutron irradiations. Such huge discrepancy is attributed to an acceleration effect of IEL on the dissipation of rapidly diffused self-interstitials and dissociation rate of defect clusters in Si.

**II. EXPERIMENTAL SETUP**

In Ref. [9], the individual 1.3MeV protons were provided by a tandem accelerator; the total proton fluence was $5×10^9$ cm$^{-2}$ and the flux was changed from $10^7$ cm$^{-2}$·s$^{-1}$ to $10^{10}$ cm$^{-2}$·s$^{-1}$. To investigate the flux dependence of dynamic annealing in absence of charge carriers, neutrons are used instead of protons. This is because 1MeV neutrons have much smaller IEL/NIEL ratio (~0.84), with IEL of $1.67×10^{-3}$ MeV cm$^2$/g and NIEL of $2.0×10^{-3}$ MeV cm$^2$/g [14]. In this configuration, the possible influence of charge carriers can be reduced to the lowest level. Neutron irradiations were performed at the Chinese Fast Burst Reactor-II (CFBR-II) of Institute of Nuclear Physics and Chemistry, China Academy of Engineering

Physics. The energy spectra of neutrons is shown in Table.1 [29], which provides an equivalent energy of 1MeV that is similar to the protons of the previous experiments. The total neutron fluence is adopted as $1\times10^{10}$ cm$^{-2}$, which is also similar to the fluence used in previous experiments of protons and other ions bombardments. While the radiation experiments are operating, the neutron flux is changed from $2.5\times10^{5}$ cm$^{-2}\cdot$s$^{-1}$ to $1\times10^{7}$ cm$^{-2}\cdot$s$^{-1}$. All of these neutron irradiations are carried out at room temperature. When the neutron fluence is $1\times10^{10}$ cm$^{-2}$, the accompanied gamma dose is less than 1rad (1 rad = 100 erg/g, 1 erg=$10^{-7}$ J). In this case, the effect of accompanied gamma ray could be neglected.

TABLE.1 Energy spectra of neutrons at CFBR-II

| Neutron energy (MeV) | Neutron spectra (%) |
|---|---|
| 0.00~0.01 | 1.5 |
| 0.01~0.02 | 0.93 |
| 0.02~0.05 | 2.51 |
| 0.05~0.10 | 3.91 |
| 0.1~0.2 | 8.3 |
| 0.2~0.5 | 24.8 |
| 0.5~1.0 | 28.8 |
| 1.0~2.0 | 19 |
| 2.0~3.0 | 5.6 |
| 3.0~4.0 | 2.26 |
| 4.0~5.0 | 1.01 |
| 5.0~6.4 | 0.62 |
| 6.4~8.2 | 0.17 |
| 8.2~10.0 | 0.07 |
| 10.0~18.0 | 0.1 |

Types of the induced defect in n-type silicon is characterized by the widely-used deep level transient spectroscopy (DLTS) technology [30, 31]. p$^{+}$-n diodes were manufactured by the same process flow as the PNP transistors. Defects of p$^{+}$-n diodes were measured using PhysTech Fourier Transform DLTS system equipped with a liquid nitrogen cryostat. The capacitance transients (DC) were measured with a 1 MHz DLTS spectrometer operated with a temperature scan from 77K to 300K. The applied reverse bias voltage and filling pulse voltage were –10 V and –0.5 V, respectively. Data presented here were taken with a rate window of 204.8 ms$^{-1}$ and the filling pulse width $t_p$ of 10 ms. The results are present in Fig. 1, from which we found that defects cannot be detected until the neutron fluence exceeds $1\times10^{12}$cm$^{-2}$.

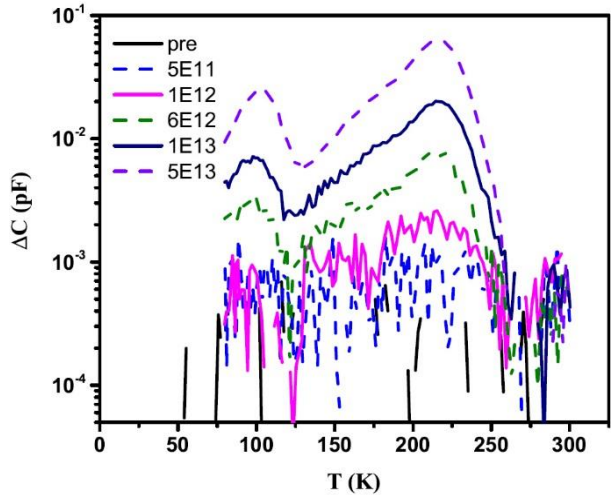

FIG. 1. DLTS spectra of p$^{+}$-n diodes bombarded by neutrons with different fluence as indicated in the figure (in unit of cm$^{-2}$).

To detect the defects generated by $1\times10^{10}$ cm$^{-2}$ fluence neutrons, recombination base currents of bipolar transistors are used. To enlarge the universality of the results, commercial operational amplifier (Op-amp) including LM324N and LM124J with PNP input stage transistor are also selected as research objects. These Op-amps are used because their input stages are very straightforward and the input bias current is directly related to the concentration of defects in the n-type silicon of the input-stage PNP transistors [32, 33].

For Op-amps, the input bias currents ($I_{ib}$) are measured by Simi3193 discrete semiconductor testing systems. For GLPNP BJTs, the excess base currents ($I_b$) are measure by Keithley 4200. PNP transistors were separated into 3 splits (3 samples in 1 split) and bombarded at fluxes of $5\times10^5$ cm$^{-2}\cdot$s$^{-1}$, $1\times10^6$ cm$^{-2}\cdot$s$^{-1}$, and $2.5\times10^6$ cm$^{-2}\cdot$s$^{-1}$, respectively. During the measurements, the emitter is grounded while the collector and base voltage is scanned from 0 V to –1 V. The base current is recorded at –0.6V. LM324N and LM124J were separated into 4 splits (16 samples in 1 split) and bombarded at fluxes of $2.5\times10^5$ cm$^{-2}\cdot$s$^{-1}$, $7.5\times10^5$ cm$^{-2}\cdot$s$^{-1}$, $1\times10^6$ cm$^{-2}\cdot$s$^{-1}$, $1\times10^7$ cm$^{-2}\cdot$s$^{-1}$, respectively. During the irradiations, all pins are shorted and grounded. During the measurements, both Op-amps are placed in an open loop configuration. The input bias current and the base current are tested at room temperature within 1~2 hours after the neutron bombardment has been finished. As indicated in Ref. [34], during this period of time, the relative slow annealing of damage can be neglected and the remaining displacement damage is often referred to as "permanent damage".

### III. RESULTS AND DISCUSSION
#### A. Flux dependence of recombination currents in bipolar devices

At a low fluence, the generated defects are very few; accordingly, the changes of $I_{ib}$ may be overwhelmed by the sample-to-sample variability [35, 36]. To avoid this confusion, the discrete results of each sample is analyzed instead of the average of the results. The results of LM324N are shown in Fig. 2 (a) as a function of the device index. The devices are ordered by the pre-irradiation currents.

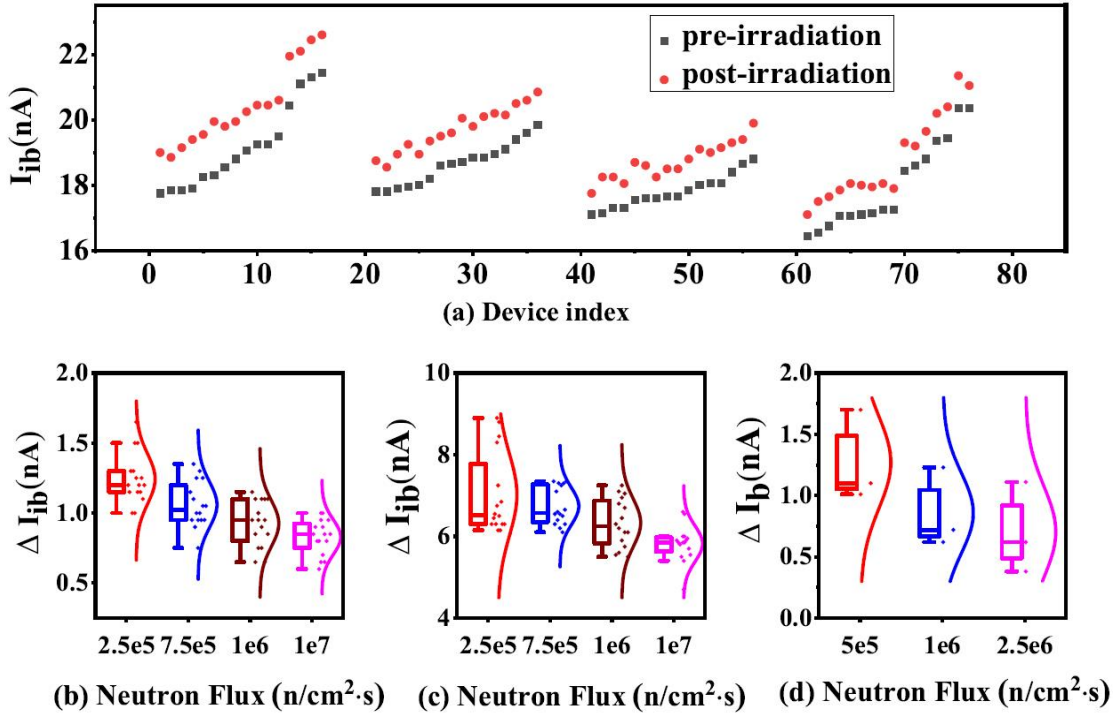

FIG. 2. (a) Pre- and post-irradiation input bias current of LM324N as a function of device index, which is ordered that the pre-irradiation value increases. (b) and (c) Box plot of the increase of input bias current of LM324N and LM124J at different neutron

flux. (d) Box plot of the increase of base current of GLPNP. During the measurements of the input bias current of LM324N and LM124J, both Op-amps are placed in an open loop configuration. During the measurements of the base current of GLPNP, the emitter is grounded while the collector and base voltage is scanned from 0 V to –1 V. The base current is recorded at –0.6V.

The information obtained from Fig. 2 (a) is as follows. First, for each flux condition the pre-irradiation input bias currents distribute randomly around 20 nA within a range of 3~5 nA, which can come from many sources [35]. Secondly, for all samples in each split, the input bias currents become larger after neutron bombardments. Although the initial values are different, the responses to a certain fluence are similar. To gain a clearer description, we display the change of input bias currents for each flux in Fig. 2(b). The data is plotted using the box view. It is clear that the $1\times10^{10}$ cm$^{-2}$ fluence neutron induces only a very small increase of the base current, about 1nA. The increase of current show a relatively big sample to sample variation, about 0.5 nA. These facts imply that generated defects in n-type silicon are very few and the concentration are very different for different samples. Remarkably, the data show a clear flux trend. Although there is an overlap between adjacent fluxes, the median for each flux clearly increases with decreasing flux. At the flux of $1\times10^7$ cm$^{-2}$·s$^{-1}$, the median of $\triangle I_{ib}$ of LM324N equals to 0.85nA. While at the flux of $2.5\times10^5$ cm$^{-2}$·s$^{-1}$, this value becomes 1.2nA.

Is the observed rule general? To answer this question, the same analysis has been made on another bipolar circuit LM124J. The results are shown in Fig. 2 (c). It is seen that, although the increment of currents is very different from those of LM324N, they also display a clear flux trend. Comparing with the LM324N case, both the medians and variations of the current increments are much larger, about 6 nA and 2 nA, respectively. When the flux decreases from $1\times10^7$ cm$^{-2}$·s$^{-1}$ to $2.5\times10^5$ cm$^{-2}$·s$^{-1}$, the median of $\triangle I_{ib}$ increases from 5.86 nA to 6.52 nA. To exclude possible influence of circuit effect, the flux dependent behavior of the base currents of a lateral PNP transistor is further investigated. The obtained data are demonstrated in Fig. 2 (d), which are found to show the same flux trend as those in Fig. 2 (b) and 2 (c).

**B. Flux dependence of displacement defects in n-type silicon**

The thin oxide layer in the PNP transistor is almost transparent for neutrons. As a result, there is almost no irradiation-induced interface states. The increment of the base current stems from the generation of displacement defects in bulk silicon and the consequent decrease of the lifetime of minority carrier ($\tau$) in the neutral base region [37]

$$\Delta I_B \propto \sum \tau_j^{-1} \propto \sum \sigma_j v_{th} N_j, \quad (1)$$

where $j$ represents the species of the defects, $\sigma_j$ is the capture cross-section of minority carriers, $N_j$ is the concentration of the defects, and $v_{th}$ is the thermal velocity of minority carriers. For proton bombardment of n-type silicon [9], five peaks of activation enthalpies of $E_1$=0.17eV, $E_2$=0.23eV, $E_3$=0.32eV, $E_4$=0.43eV, and $E_5$=0.45eV are found in the DLTS spectra, from which defects of $VO$ ($E_1$), $V_2^{2-}$ ($E_2$), $V_2^{-}$ ($E_4$), and hydrogen related defects ($E_{3,5}$) are identified [31, 38–44].

As obtained in Fig. 1, for neutron bombardment, the DLTS spectra of p$^+$-n diodes manufactured with the same processes of PNP transistors show $E_1$ ($VO$ defect) and $E_4$ ($V_2^{-}$ defect), where the shoulder to the low-temperature side of the $E_4$ peak was assumed to a multi-vacancy related defect, such as $V_3$ [45,46]. For neutron irradiation, $E_2$ is not detected because it is suppressed by $E_4$ [47, 48]; the hydrogen related defects ($E_{3,5}$) are induced by proton [44] hence cannot be obtained by neutron irradiations. The base current is mainly contributed by the defects that produce energy levels in the middle third of the Si band gap [38]. Here the energy level of $V_2^-$ lies in the middle third of the Si band gap. In Eq. (1) neither $\sigma_j$ of $V_2^-$ nor $v_{th}$ changes during the bombardment, therefore the measured flux dependent behavior of $\triangle I_B$ reflects directly the flux dependence of the concentration of $V_2^-$ defects, which further indicates a positive flux dependence of dynamic annealing. This effect is the same as the results obtained from proton and other ions bombardments [9–13]. From this fact, we can answer the question at the very beginning of the paper: the positive flux dependence of dynamic annealing also arises when the IEL is strongly suppressed (by using neutron).

However, there is a big discrepancy between the sensitive flux regions of the bombardments of neutron and other ions. As observed in Refs. [9] and [13], the sensitive flux for proton bombardment is $1\times10^{10}$ cm$^{-2}$ s$^{-1}$ and becomes smaller for ions of heavier mass or larger energy. The reason is that the sensitive flux range is determined by the mass of the ions [13]. Based on this rule, neutron and proton implantations should have similar sensitive fluxes, because the mass of neutron and proton are almost the same [13]. However, as observed in Fig. 2 the sensitive flux range for neutron is $\sim1\times10^{6}$ cm$^{-2}\cdot$s$^{-1}$. This is much lower than the sensitive flux of proton bombardment and even smaller than that of $^{120}$Sn bombardment ($\sim2\times10^{8}$ cm$^{-2}\cdot$s$^{-1}$) with the biggest mass. The obvious difference up to 10,000 times of the sensitive fluxes between neutron and proton violates the rule.

**C. Mechanisms of neutron's flux effect and discrepancy in sensitive fluxes**

We believe that the flux effects of neutron irradiations stem from the similar origin as the ion implantation experiments. In Refs. [9, 13] the effect has been attributed to the dissipation of rapidly diffused silicon interstitials, which influences the inter-cascade recombination between self-interstitials and vacancies.

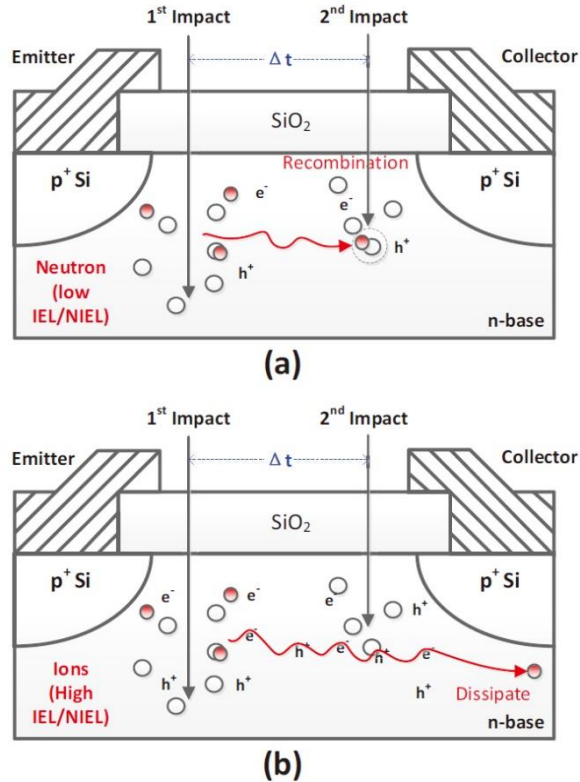

FIG. 3. Schematic show of the difference of the annealing processes between (a) neutron and (b) proton bombardments.

After a damage cascade is generated, the vacancies are forming more stable vacancy-related defects ($V_2^-$, $VO$, $V_2^{2-}$, etc.), while the mobile self-interstitials would leave the cascade and rapidly diffuse through the sample, see Fig. 3 (a). $\triangle t$ represents the time intervals between the successive impacts. For high enough flux, a subsequent damage cascade can be generated before a large portion of interstitials have dissipated out of the systems. As a result, the diffusing self-interstitials would have decent possibility to recombine with the vacancy-related defects of the second cascade, see Fig. 3 (a). In contrast, at low flux, most self-interstitials have dissipated before the subsequent cascade is generated and the inter-cascade recombination would be weak. Therefore, the higher flux bombardment would lead to an enhanced annealing of the vacancy-related defects which leads to the above observed reduction of defects buildup.

The huge difference in the sensitive flux of neutron and protons can be explained based on this picture. The picture shows that the diffusion speed of self-interstitials is a characteristic parameter limiting the sensitive flux. For proton bombardment, the self-interstitials diffuse much faster than those for neutron bombardment, due to the presence of IEL induced charge carriers [24–28], see Fig.3 (b). In presence of the carriers, the quasi Fermi energy in the sample is shifted and the defects can change from one charge state to another. As suggested by *ab-initio* calculations [26–28], the self-interstitials of different charge states have different diffusion barriers. For example, the self-interstitials $I^{++}$ at its equilibrium site $H$ can change to another charge state $I^+(H)$ that is not at equilibrium. The latter will easily move to its equilibrium site $I^+(B)$ without having to overcome a barrier. Therefore, the diffusion of self-interstitials is largely enhanced in presence of carriers. Accordingly, the flux effect for proton can be observed at a high flux range. For neutron irradiation, the diffusion speed of self-interstitials is much slower, which leads to the ultra-slow dynamic annealing. Hence, the flux effect can be observed at a much lower flux range. The dissociation rate of defect cluster is another characteristic parameter limiting the sensitive flux. For proton bombardment, IEL-induced charge carriers will recombine, which will enhance defect reactions in clusters [49, 50]. Accordingly, the dissociation rate increases significantly and the flux effect should be observed at a much higher flux range.

The discrepancy in sensitive fluxes may also be influenced by the different nature and spatial distribution of defects generated by neutron and proton. For 1 MeV protons irradiation, the incoming ions would collide with both silicon atoms and electrons. The defects are primarily accumulated at the end of range when it has lost most energy and the nuclear stopping becomes dominate [51]. The depth is estimated to be several μm [52]. In contrast, due to the charge neutrality, MeV neutrons will not interact with the electrons as protons and the collisions only takes place with the atomic nuclei. Therefore, neutrons have much smaller scattering cross section and larger penetration range. The generated defects are nearly uniformly distributed in the bulk silicon with much sparser distributions than protons/ions of the same fluence. The different distribution may require more time for mobile partible to diffuse and interact among defects, which could also influence the sensitive flux.

## IV. CONCLUSION

To investigate the flux dependence of dynamic annealing of displacement defects in silicon in absence of charge carriers, neutron bombardment of equivalent fluence and different fluxes have been performed on bipolar transistors and circuits, whose ultra-sensitivity recombination currents provide a probe for the concentration of very few $V_2^-$ defects in silicon. Similar to proton bombardment, the defect concentration (strength of dynamic annealing) is found to decrease (increase) with increasing neutron flux. However, the sensitive flux range (~1×10$^6$ cm$^{-2}$·s$^{-1}$) is about 4 orders of magnitude lower, despite of similar mass and energy of the two particles. The huge discrepancy is attributed to ionization energy-loss-induced large difference in diffusion velocity of Si interstitials and dissociation rate of defect clusters in a model considering the dissipation of diffusing Si interstitials. The obtained effect and unraveled mechanisms will be helpful for the study of the basic process of dynamic annealing in silicon and the role of ionization energy loss in it.

**ACKNOWLEDGMENTS**

This work was supported by the Science Challenge Project under grant no. TZ2016003-1 and NSFC under grant nos. 11804313 and 11404300.